\documentclass[twocolumn,showpacs,preprintnumbers,amsmath,prd,amssymb,superscriptaddress]{revtex4}

\pdfoutput=1

\usepackage{graphicx}
\usepackage{dcolumn}
\usepackage{bm}
\usepackage{multirow}

\def\cpkkd{\rm{kg^{-1}keV^{-1}day^{-1}}}
\def\keVee{\rm{keV_{ee}}}
\def\eVee{\rm{eV_{ee}}}
\def\mwimp{\rm{m_{\chi}}}
\def\effbs{\epsilon _{\rm BS}}
\def\lmbdbs{\lambda _{\rm BS}}
\def\pge{{\it p}{\rm Ge}}
\def\nge{{\it n}{\rm Ge}}
\def\b0{{\rm B_0}}
\def\s0{{\rm S_0}}

\begin{document}

\preprint{AS-TEXONO/13-02}

\title{
Differentiation of Bulk and Surface Events
in p-type Point-Contact Germanium Detectors
for Light WIMP Searches
}

%

\newcommand{\as}{Institute of Physics, Academia Sinica, 
Taipei 11529, Taiwan.}
\newcommand{\thu}{Department of Engineering Physics, Tsinghua University,
Beijing 100084, China.}
\newcommand{\metu}{Department of Physics,
Middle East Technical University, Ankara 06531, Turkey.}
\newcommand{\deu}{Department of Physics,
Dokuz Eyl\"{u}l University, Buca, \.{I}zmir 35160, Turkey.} 
\newcommand{\ciae}{Department of Nuclear Physics,
Institute of Atomic Energy, Beijing 102413, China.}
\newcommand{\bhu}{Department of Physics, Banaras Hindu University,
Varanasi 221005, India.}
\newcommand{\nku}{Department of Physics, Nankai University,
Tianjin 300071, China.}
\newcommand{\scu}{Department of Physics, Sichuan University,
Chengdu 610065, China.}
\newcommand{\ks}{Kuo-Sheng Nuclear Power Station,
Taiwan Power Company, Kuo-Sheng 207, Taiwan.}
\newcommand{\corr}{htwong@phys.sinica.edu.tw} 

\author{ H.B.~Li }  \affiliation{ \as }
\author{ L.~Singh }  \affiliation{ \as } \affiliation{ \bhu }
\author{ M.K.~Singh }  \affiliation{ \as } \affiliation{ \bhu }
\author{ A.K.~Soma }  \affiliation{ \as } \affiliation{ \bhu }
\author{ C.H.~Tseng }  \affiliation{ \as }
\author{ S.W.~Yang }  \affiliation{ \as }
\author{ M.~Agartioglu }  \affiliation{ \as } \affiliation{ \deu }
\author{ G.~Asryan }  \affiliation{ \as }
\author{ Y.C.~Chuang }  \affiliation{ \as }
\author{ M.~Deniz } \affiliation{ \deu }
\author{ T.R.~Huang }  \affiliation{ \as }
\author{ G.~Kiran Kumar } \affiliation{ \as }
\author{ J.~Li }  \affiliation{ \thu } 
\author{ H.Y.~Liao }  \affiliation{ \as }
\author{ F.K.~Lin }  \affiliation{ \as }
\author{ S.T.~Lin }  \affiliation{ \as } \affiliation{ \deu }
\author{ S.K.~Liu } \affiliation{ \scu }
\author{ V.~Sharma }  \affiliation{ \as } \affiliation{ \bhu }
\author{ Y.T.~Shen }  \affiliation{ \as }
\author{ V.~Singh }  \affiliation{ \bhu }
\author{ H.T.~Wong } \altaffiliation[Corresponding Author: ]{ \corr } \affiliation{ \as }
\author{ Y.C.~Wu } \affiliation{ \thu }
\author{ Y.~Xu } \affiliation{ \as } \affiliation{ \nku }
\author{ C.X.~Yu } \affiliation{ \as } \affiliation{ \nku }
\author{ Q.~Yue } \affiliation{ \thu }
\author{ W.~Zhao }  \affiliation{ \thu } 


\collaboration{TEXONO Collaboration}



\date{\today}

\begin{abstract}

The p-type point-contact germanium detectors
are novel techniques offering kg-scale
radiation sensors with sub-keV sensitivities.
They have been used for light Dark Matter WIMPs searches
and may have potential applications in neutrino physics.
There are, however, anomalous surface behaviour
which needs to be characterized and understood.
We describe the methods and results of 
a research program whose goals are
to identify the bulk and surface events 
via software pulse shape analysis techniques, 
and to devise calibration schemes to evaluate
the selection efficiency factors.
Efficiencies-corrected 
background spectra from
the low-background facility 
at Kuo-Sheng Neutrino Laboratory are derived.

\end{abstract}


\maketitle

\section{Introduction}

Searches and identification of dark matter~\cite{cdmpdg12} 
are at the forefronts of experimental research.
Germanium detectors with sub-keV sensitivities
have been demonstrated as efficient means to probe
Weakly Interacting Massive Particles 
(WIMPs, denoted by $\chi$), of mass 
$\mwimp \sim 1 - 10 ~ {\rm GeV}$~\cite{ulege,texono2007}.
This motivates development of p-type point-contact
germanium detectors ($\pge$)~\cite{ppcge}. 
This novel detector technique is also adopted in the
studies of neutrino-nucleus coherent scattering
with reactor neutrinos~\cite{ulege}.
In both cases, the interaction channel is
\begin{equation}
\chi ( \nu ) + N \rightarrow \chi ( \nu ) + N  ~~ ,
\end{equation}
where $N$ denotes the nucleus. 
The experimental signatures are the nuclear recoils,
posing the challenging requirements of low background
and low threshold to the detectors.
Allowed region together with annual modulation signatures
at $\mwimp \sim 8 ~{\rm GeV}$ and 
spin-independent elastic cross-section 
$\sigma_{\chi N} \sim 3 \times 10^{-41} ~ {\rm cm^2}$ 
were implied by the CoGeNT experiment~\cite{cogent},
while limits were derived by the 
TEXONO~\cite{texono2013} and CDEX-1~\cite{cdex12013} 
experiments. 

The surface events in $\pge$ exhibit 
anomalous behaviour~\cite{gesurface,cogent,texonopgebs,texono2013}.
It may limit the physics sensitivities, 
and can lead to false interpretation of the data.
The analysis of these anomalous surface events 
is therefore an important experimental challenge to overcome 
before the full potentials of this novel detector
technique at sub-keV energy can be realized.

We document this aspect of the 
TEXONO experiment~\cite{texono2013} in this report. 
The physics origin, rise-time measurements,
separation of bulk from surface events as well as the
derivations of efficiency factors and
the associated uncertainties 
are discussed in the subsequent sections.
The data adopted to illustrate the
analysis procedures are from a $\pge$ 
whose target is a cylindrical germanium crystal 
of 60.1~mm in diameter and 60.8~mm in height. 
The data size is 39.5~kg-days~\cite{texono2013}.
The application of the same analysis on
other $\pge$ detectors or to different
data set from the same detector
gives consistent behaviour and results.

\section{Experiment Overview}

Signals from the point-contact of the $\pge$
are supplied to a reset preamplifier.
The output is distributed to
(a) a fast timing amplifier (TA) digitized at 200~MHz
which keeps the rise-time information, and 
(b) a shaping amplifier (SA) at 6~$\mu$s shaping time 
digitized at 60~MHz which provides the 
trigger and measurement of energy (denoted by $T$). 
The pedestal fluctuation RMS is 55~$\eVee$, 
the test pulser FWHM is 141~$\eVee$
while the electronic noise edge is at
400~$\eVee$ $-$
electron-equivalent energy is used
throughout in this article to denote detector response.
The analysis threshold for this work is 500~$\eVee$.

\begin{figure}
\includegraphics[width=8.5cm]{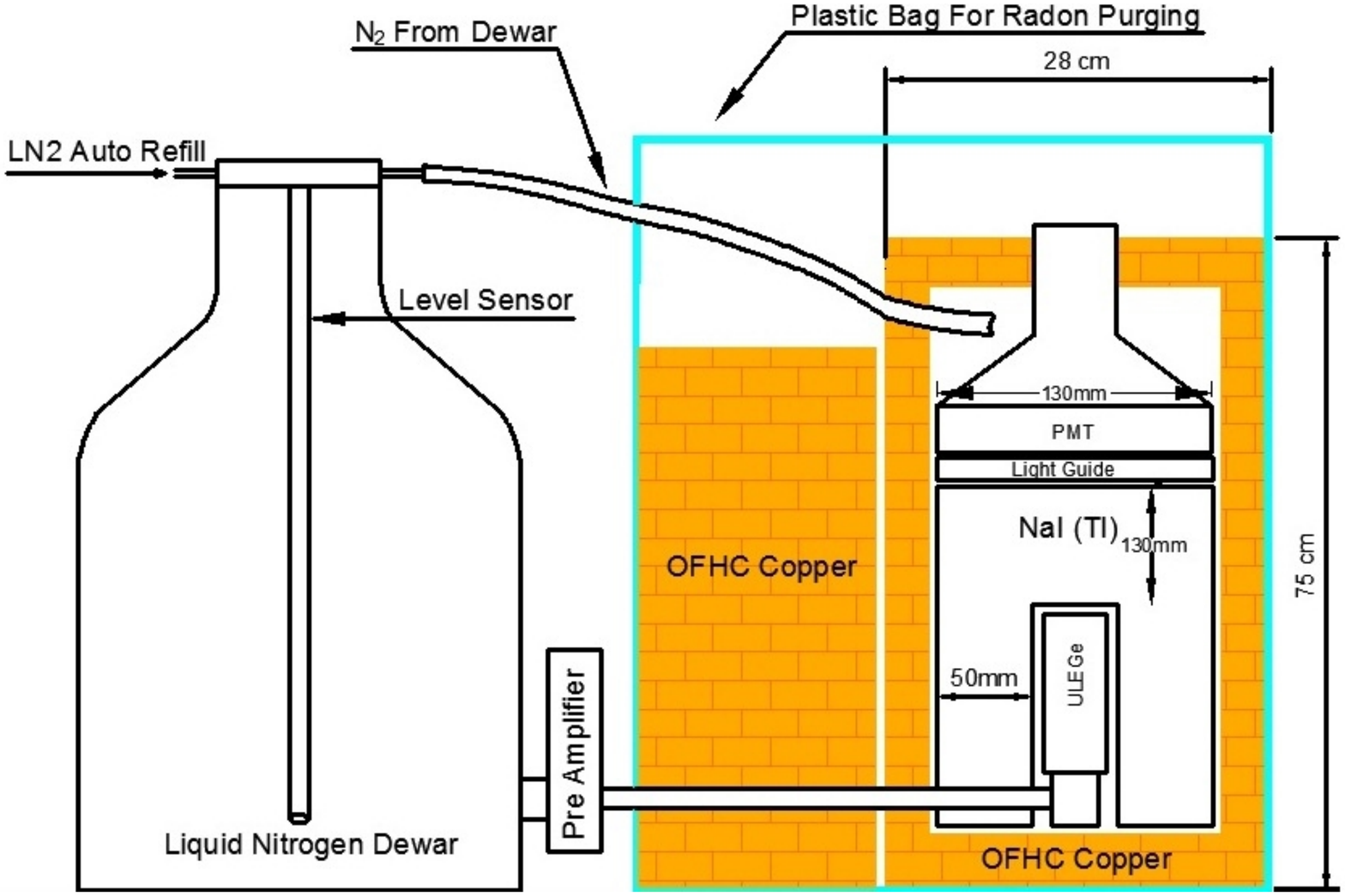}
\caption{
Schematic diagram of the experimental set-up which
includes the $\pge$ and NaI(Tl) scintillator.
The hardware is placed inside a 50-ton
shielding structure, surrounded by plastic
scintillators as cosmic-ray vetos.
}
\label{fig::setup}
\end{figure}

The detector system is installed at the
Kuo-Sheng Reactor Neutrino Laboratory (KSNL)
for the studies of neutrino physics~\cite{ksnlneutrino} 
and light WIMP searches~\cite{texono2013}.
The schematic diagram of the experimental set-up
is given in Figure~\ref{fig::setup}.
Software pulse shape analysis identifies
electronic noise background from the
physics events characterized by 
genuine creation of electron-hole pairs
in the crystal.
The NaI(Tl) crystal scintillator serves as 
anti-Compton (AC) detector while the surrounding
plastic scintillator panels provide 
the cosmic-ray (CR) veto.
The physics events are categorized by
``AC$^{-(+)}$$\otimes$CR$^{-(+)}$'', where
the superscript $-(+)$
denotes anti-coincidence(coincidence) 
with the $\pge$ signals.
The selection procedures as well as the
derivation of their efficiencies 
have been well-established~\cite{ksnlneutrino,texono2007}
through the several experiments conducted at KSNL
with this baseline design.

\begin{figure}
{\bf (a)}\\
\includegraphics[width=8.5cm]{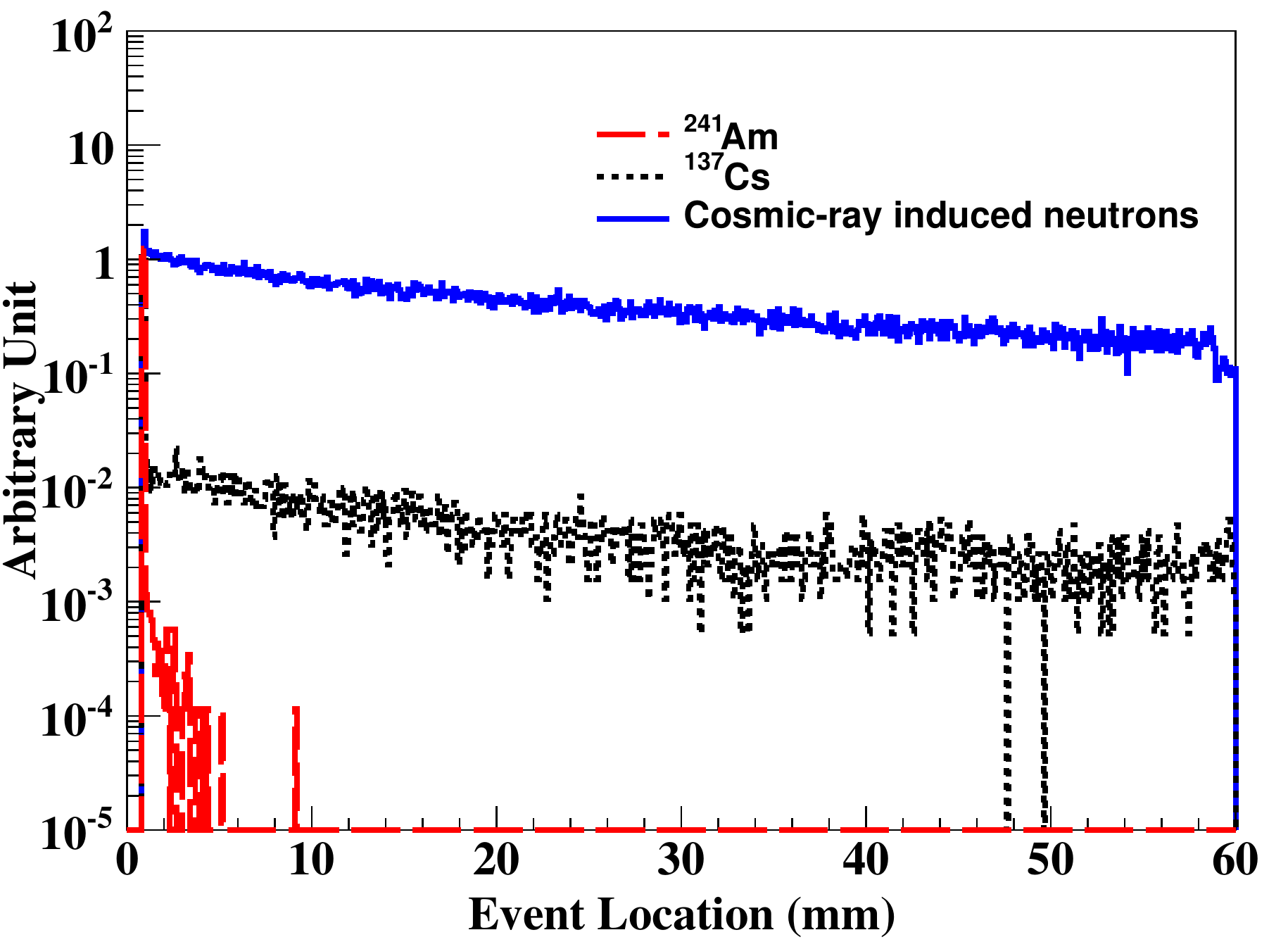}
{\bf (b)}\\
\includegraphics[width=8.5cm]{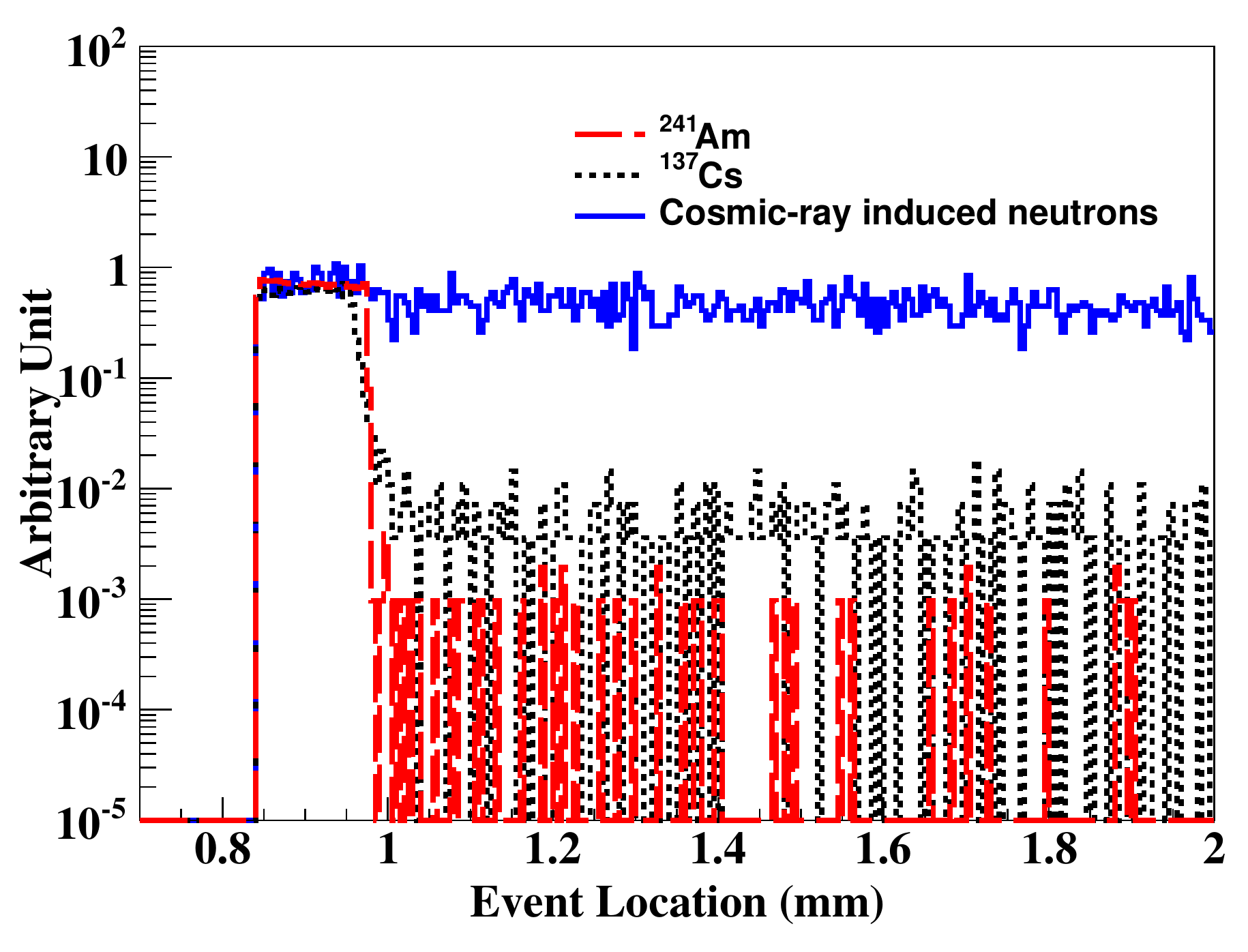}
\caption{
(a) The simulated penetration-depth distributions of various samples: 
low and high energy $\gamma$-rays in 
$^{241}$Am and $^{137}$Cs, respectively, 
as well as cosmic-ray induced high energy neutrons, where
energy depositions are less than 6~$\keVee$
after folding in the surface quenching effects.
The different distributions are normalized by their event
numbers at the first bin.
(b) Features at surface are elaborated. 
}
\label{fig::eventdepth}
\end{figure}

The tagging-manifolds correspond to
events from different physics origins.
Nuclear recoil $( \chi / \nu ) N$ events 
are uncorrelated with other detector components and uniformly
distributed in the $\pge$ volume. 
Therefore, the candidate events 
are tagged by AC$^{-}$$\otimes$CR$^{-}$, 
while AC$^{+}$$\otimes$CR$^{-}$
and AC$^{-}$$\otimes$CR$^{+}$ select 
ambient gamma and 
cosmic-ray induced high energy neutron events,
respectively.
In addition, calibration data are taken with
low and high energy $\gamma$-sources
($^{241}$Am at 59.5~$\keVee$ and $^{137}$Cs at 661.7~$\keVee$,
respectively).
These sources give rise to 
events with different penetration-depth 
distributions, as depicted in 
Figures~\ref{fig::eventdepth}a\&b.
They therefore play complementary roles
in probing the detector response over
the entire active volume.

\begin{figure}
\includegraphics[width=8.5cm]{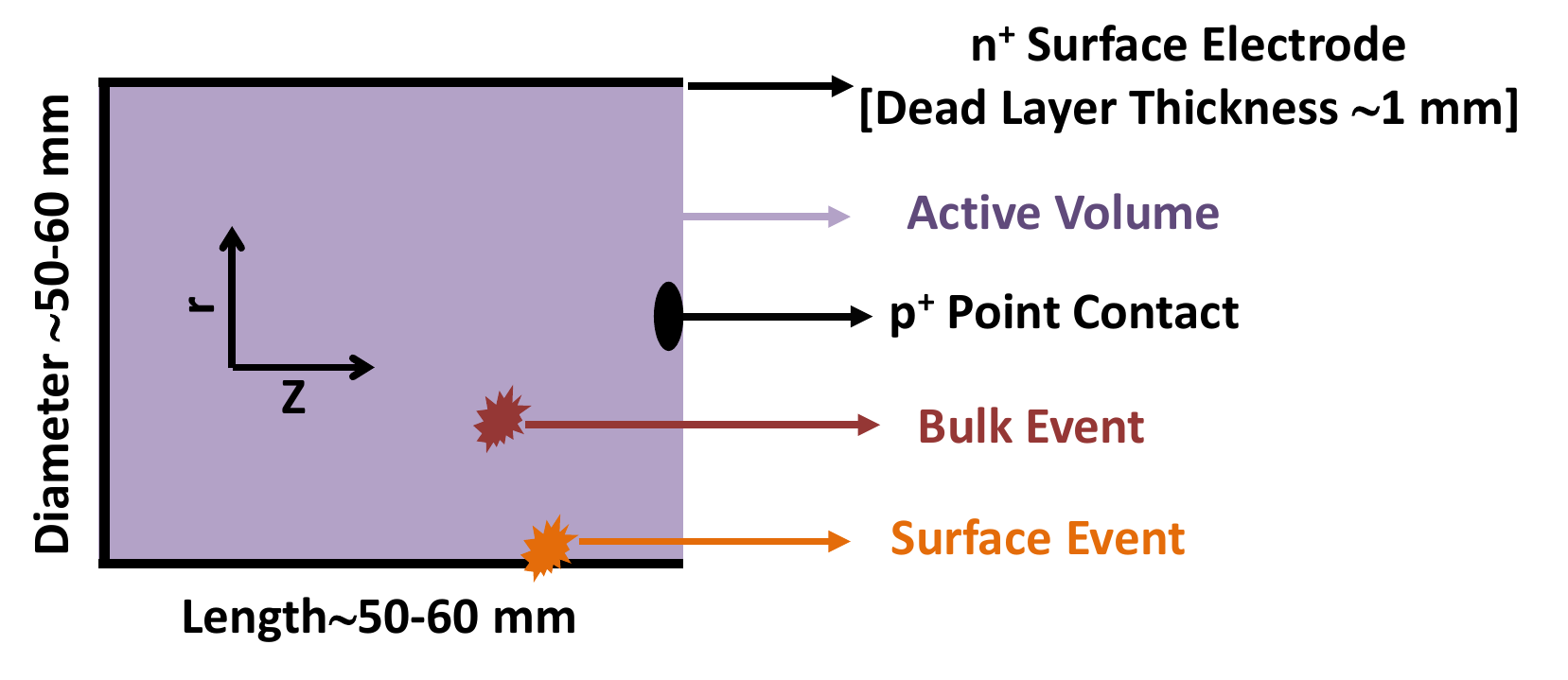}
\caption{
Schematic diagram of the Ge-crystal in $\pge$,
showing the central point-contact and surface electrodes.
}
\label{fig::ppcge}
\end{figure}


\section{Bulk and Surface Events}

\subsection{Physics Origin}

The schematic diagram of a typical $\pge$ sensor is
displayed in Figure~\ref{fig::ppcge}. 
The crystal is made of p-type germanium.
The outer surface electrode is at positive high voltage 
towards which the electrons are drifted.
The small central contact electrode 
is at zero-potential from which electrical 
signals are extracted. 

\begin{figure}
{\bf (a)}\\
\includegraphics[width=8.5cm]{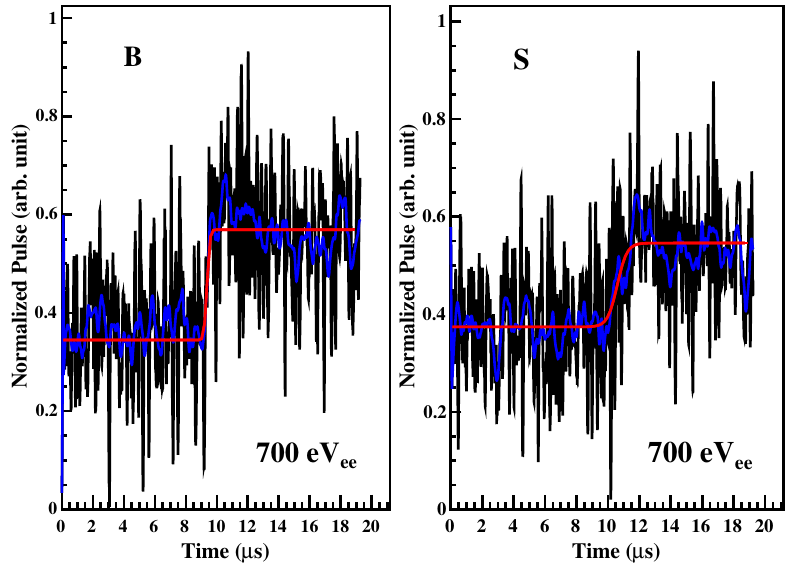}
{\bf (b)}\\
\includegraphics[width=8.5cm]{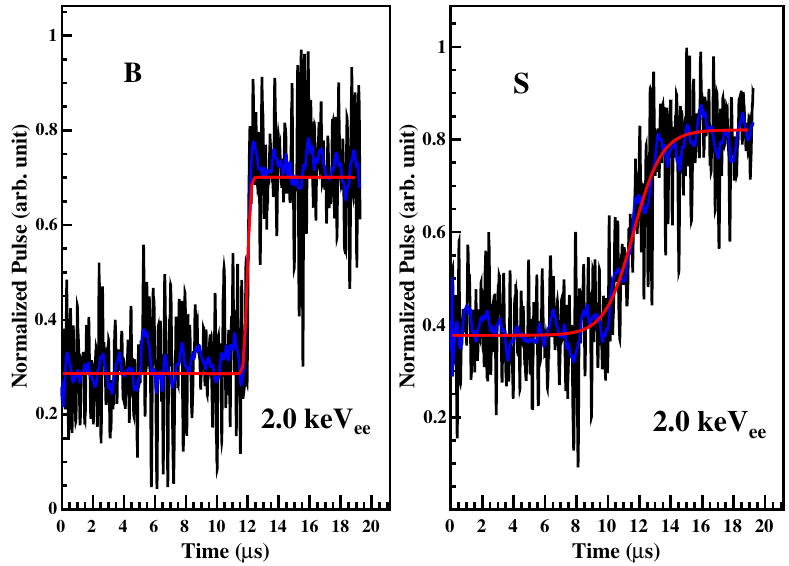}
\caption{
Typical B/S events at (a) 700~$\eVee$ and (b) 2~$\keVee$ energy,
showing the raw (black) and smoothed (blue) pulses, together with the
best-fit functions (red).
}
\label{fig::bsevents}
\end{figure}

The outer surface electrode is fabricated by 
lithium diffusion.
It has a finite thickness of typically $\sim$1~mm.
Electron-hole pairs produced by radiations at the
surface (S) layer are subjected to a weaker drift field
than those at the crystal bulk region (B).
A portion of the pairs will recombine while the 
residual will induce signals which are weaker 
and slower than those originated in B.
That is, the S-events have only partial 
charge collection and slower rise-time.
The thickness of the S layer for 
the $\pge$ in this work
is measured to be (1.16$\pm$0.09)~mm,
via the comparison of
simulated and observed intensity ratios
of $\gamma$-peaks from
a $^{133}$Ba source~\cite{ba133}.
In addition, through the comparison
of the B- and S-intensities and spectra
with the various $\gamma$-sources,
it can be derived that there is no 
charge collection at a depth of less than 0.84~mm.
The dead and inactive layers are 
illustrated in the event location distributions
of Figures~\ref{fig::eventdepth}a\&b.
Only events with complete charge
collection are considered as bulk events.
The corresponding fiducial mass 
for the B-region is 840~g.

\begin{figure}
{\bf (a)}\\
\includegraphics[width=8.5cm]{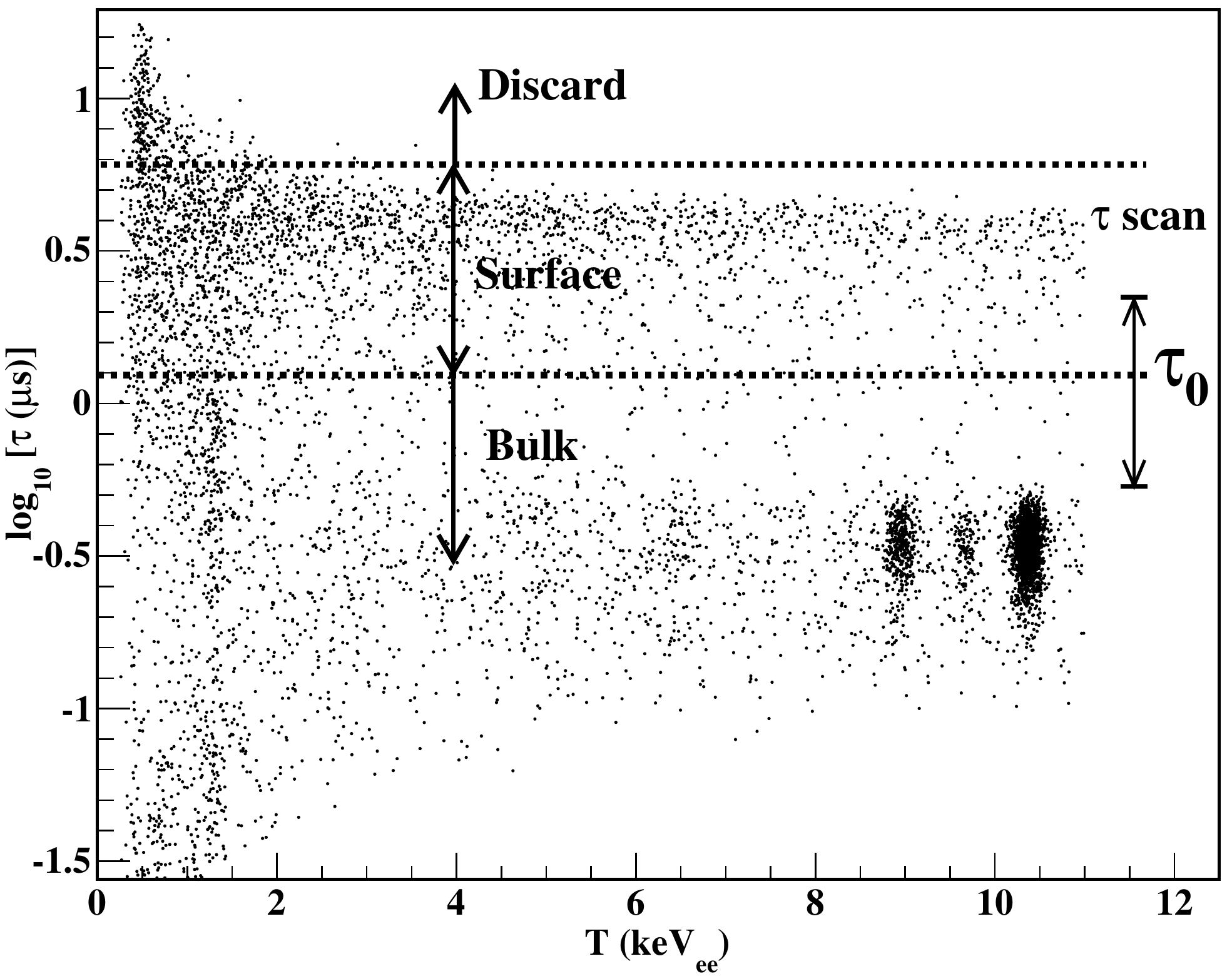}
{\bf (b)}\\
\includegraphics[width=8.5cm]{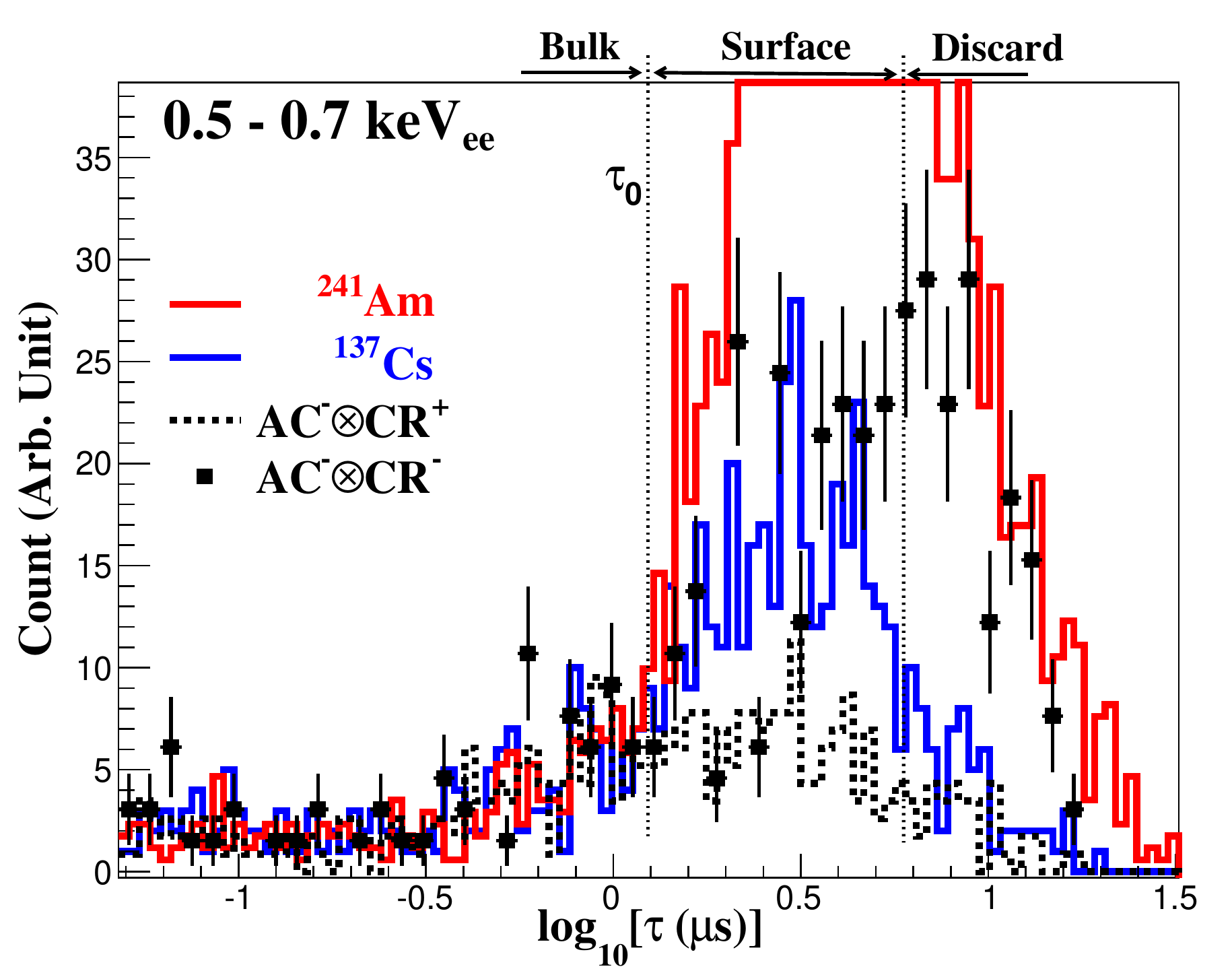}
\caption{
(a)
The $\tau$ versus $T$ scatter plot for
the AC$^{-}$$\otimes$CR$^{-}$ tags
which select $( \chi / \nu ) N$ candidate
events.
(b)
The $\tau$ distribution at 700~$\eVee$, 
comparing the candidate events with those 
of $^{241}$Am, $^{137}$Cs,
and cosmic-induced neutrons (AC$^{-}$$\otimes$CR$^{+}$).
The various histograms are normalized by the areas
of the bulk band.
}
\label{fig::bscut}
\end{figure}


\begin{table*}
\begin{ruledtabular}
\begin{tabular}{lccccccc}
& \multicolumn{3}{c}{Bulk-Band} &   & \multicolumn{3}{c}{Surface-Band} \\
& LE & ME & HE & $\mid$ & LE & ME & HE \\ \hline
$< \tau >$  ($\mu$s) &
0.52  & 0.31 & 0.35  & $\mid$ & 4.9 & 4.3 & 4.1 \\
$\sigma_{fit}$ ($\mu$s) & 0.025 & 0.023 & 0.009 & $\mid$ 
& 0.20 & 0.05 & 0.02 \\
$\sigma_{i}$ ($\mu$s) &
\multicolumn{3}{c}{0.065} & $\mid$ & \multicolumn{3}{c}{0.66} \\ 
$\sigma_{m}$ ($\mu$s) & 
1.21 & 0.28 & 0.11 & $\mid$ & 2.75 & 1.24 & 0.69 \\
$\sigma_{\tau}$ ($\mu$s) & 
1.21 & 0.27 & 0.09  & $\mid$ & 2.67 & 1.05 & 0.21 \\
$ | < \tau > - ~ \tau_{0} ~ |  ~ / ~  \sigma_{m} $ & 
0.6 & 3.3 & 8.0 & $\mid$ & 1.3 & 2.5 & 4.2  \\
\end{tabular}
\end{ruledtabular}
\caption{
Summary of the rise-time ($\tau$) measurements
for bulk (B) and surface (S) bands at 
low (LE: 500$-$700~$\eVee$),
medium (ME: 1.5$-$2~$\keVee$)
and high (HE: 6$-$8~$\keVee$) energy.
All width and resolution values correspond to the RMS of the distributions
with the mean-values at $< \tau >$.
The fitting errors ($\sigma_{fit}$)
correspond to those due to the analysis algorithms.
The intrinsic width ($\sigma_{i}$) is defined by the
surface-dominating $^{241}$Am events at 10~$\keVee$ and 
by the homogeneously-distributed
Ga-X-rays at 10.37~$\keVee$ for the S- and B-bands, respectively.
Combining $\sigma_{i}$ and the $\tau$-resolution ($\sigma_{\tau}$)
in quadrature gives the measured width of the band ($\sigma_{m}$).
The optimal $\tau$-cut ($\tau_0$) is set at 1.23~$\mu$s.
The last row characterizes the separation of $< \tau >$
from $\tau_0$, in unit of $\sigma_{m}$.
}
\label{tab::tauband}
\end{table*}


This anomalous surface charge collection effect 
has been studied in early literature~\cite{gesurface}.
However, the quenched S-signals are mostly
$\alt$1~$\keVee$, 
below the typical Ge detector threshold of
a few $\keVee$. Consequently, 
the S-layer in p-type germanium detectors 
were mostly classified as ``dead''.
With the advent of $\pge$ and 
the physics region of interest moving to the
sub-keV range, it was observed~\cite{cogent,texono2013,texonopgebs}
that these anomalous S-events do exist and 
would dominate the low energy background.
The identification of the S-events 
and the knowledge of efficiency factors 
therefore become crucial to fully exploit
the sub-keV sensitivities of $\pge$.

\subsection{Rise-time Measurement}

Typical TA-signals for B/S-events at low ($\sim$700~$\eVee$)
and high ($\sim$2~$\keVee$) energy are depicted in 
Figures~\ref{fig::bsevents}a\&b, respectively.
The TA rise-time ($\tau$) is parametrized by 
the hyperbolic tangent function 
\begin{equation}
\frac{1}{2}~{\rm A_0}
\times ~ \tanh(\frac{t-{\rm t_0}}{\tau}) ~ + ~ {\rm P_0} ~~ ,
\label{eq::taufct}
\end{equation}
where $\rm{A_0}$, ${\rm P_0}$ and $\rm{t_0}$ are, respectively,
the amplitude, pedestal offset and timing offset.
The values of ${\rm P_0}$ and $\rm{A_0}$ are 
evaluated from the TA-pulses prior to the transition edge
and through the difference of the asymptotic levels, respectively.
The time difference as a function of energy 
between the TA-edge and the DAQ-trigger instant 
defined by the SA signals is pre-determined, 
and provides constraints on $\rm{t_0}$.
The raw TA-pulses are first smoothed by
the Softies-Kola filter~\cite{sgfilter} 
and fitted to Eq.~\ref{eq::taufct} with ($\tau$,$\rm{t_0}$)
as free parameters.
The results are then adopted as initial values
to another fit of the same function
directly on the raw pulse. 
The two procedures are complementary $-$ 
$\rm{t_0}$ and $\tau$ are sensitive to 
the smoothed and raw pulses, respectively.
The smoothed and best-fit functions are overlaid 
to the raw FADC signals in Figure~\ref{fig::bsevents}.

A small fraction ($<$8\%) of events at low energy
fails the fitting procedures.
These events are excluded for subsequent analysis.
The signal efficiency is accounted for
through the survival probability of 
the doubly-tagged AC$^{+}$$\otimes$CR$^{+}$ samples,
which is 80\% at ~500~$\eVee$.

The scatter plot of $\tau$ versus $T$
for the AC$^{-}$$\otimes$CR$^{-}$ 
events at KSNL is displayed in
Figure~\ref{fig::bscut}a.
Events with $\tau$ less(greater) than 
a selected cut-value $\tau_0$ 
(=1.23~$\mu$s in this analysis)
are categorized as B(S).
A summary of the width and resolution contributions
to the bands is given in Table~\ref{tab::tauband}.
The fitting errors ($\sigma_{fit}$) correspond to those
due to the software algorithms. 
They are small compared to the measured width ($\sigma_{m}$)
from the various calibration data set.
There are two contributions to the $\tau$-width:
(i) the intrinsic width $\sigma_{i}$ is due to
the non-uniform response over the detector volume 
producing different $\tau$,
while
(ii) the $\tau$-resolution $\sigma_{\tau}$ is due to
fluctuations of pulse shape
for events at the same nominal $\tau$.
The $\sigma_{i}$'s are measured 
from the surface-dominating $^{241}$Am events
at 10~$\keVee$ and the homogeneously-distributed
Ga-X-rays at 10.37~$\keVee$ for the S- and B-bands, 
respectively, while $\sigma_{\tau}$ is 
derived via $\sigma_{\tau}^2 = \sigma_{m}^2 - \sigma_{i}^2$.

At $T > 1.5~\keVee$,
the $\sigma_{m}$ is much less than 
the separation of the bands from $\tau_0$.
The measurements of $\tau$ therefore
provide valid information on the locations of 
the events and, in particular, can efficiently
differentiate the S- from B-events.
This behaviour manifest as a distinct
two-band structure in Figure~\ref{fig::bscut}a,
with a small fraction 
(about 8\% within 3$-$6~$\keVee$ in the 
AC$^-$$\otimes$CR$^-$ sample)
of events in the intermediate transition zone.
By studying the corresponding fractions of events 
with $^{241}$Am ($<$1\%) and $^{137}$Cs (7.5\%)
$\gamma$-sources, a thickness of 0.16~mm 
for this zone is derived.
The choice of $\tau_0$ is equivalent to a definition
of the spatial borderline between B/S
within this transition thickness. 
This gives rise to a systematic uncertainty 
in the evaluation of the $\pge$ fiducial mass.
It translates to about 3\% of the total error 
at 500~$\eVee$ which, as displayed in Table~\ref{tab::error},
is negligible compared to the other error sources.

At $T < 1.5~\keVee$
where $\sigma_{m}$ is comparable
to the band separation, there exist
contaminations between the B- and S-events
which lead to the merging of the bands.
The methods and results to evaluate the leakage factors 
and to correct the measured spectra
are discussed in the subsequent sections.

The $\tau$-distribution of the candidate samples 
at 700~$\eVee$ 
are displayed in Figure~\ref{fig::bscut}b, 
together with those
due to low and high energy $\gamma$-rays
and high energy neutrons.
The B-events near analysis threshold have similar distributions
for all samples and are independent of locations. 
Different distributions in the S-events
are observed. These can be accounted for 
by their different penetration profiles 
depicted in Figure~\ref{fig::eventdepth}b.

\begin{figure}
{\bf (a)}\\
\includegraphics[width=8.5cm]{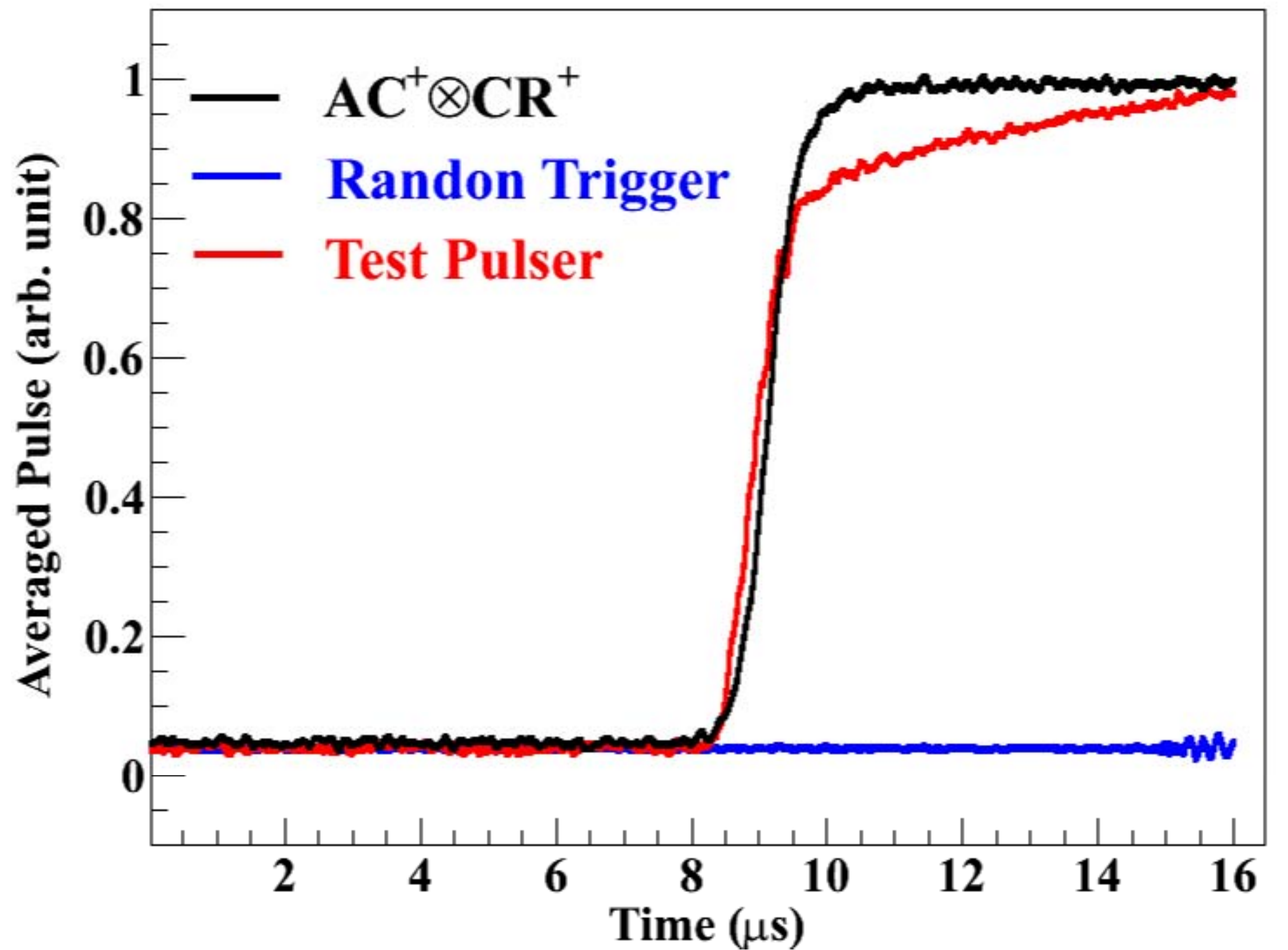}
{\bf (b)}\\
\includegraphics[width=8.5cm]{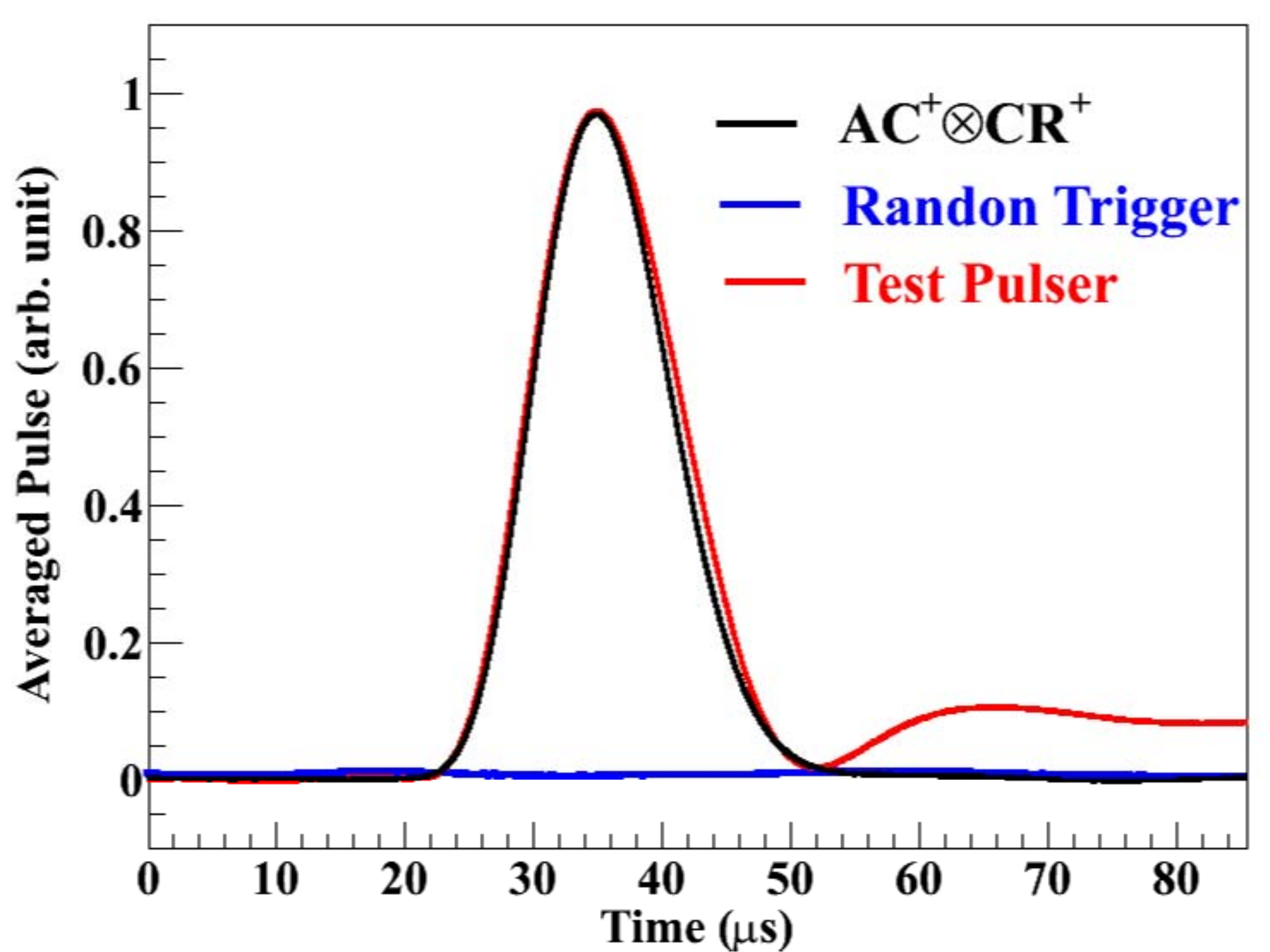}
\caption{
Averaged shapes of the (a) fast timing and
(b) shaped pulses of events due to
random trigger, test pulser and
physics-samples with
the AC$^{+}$$\otimes$CR$^{+}$ tag.
}
\label{fig::pulseshape}
\end{figure}

\begin{figure}
{\bf (a)}\\
\includegraphics[width=7.5cm]{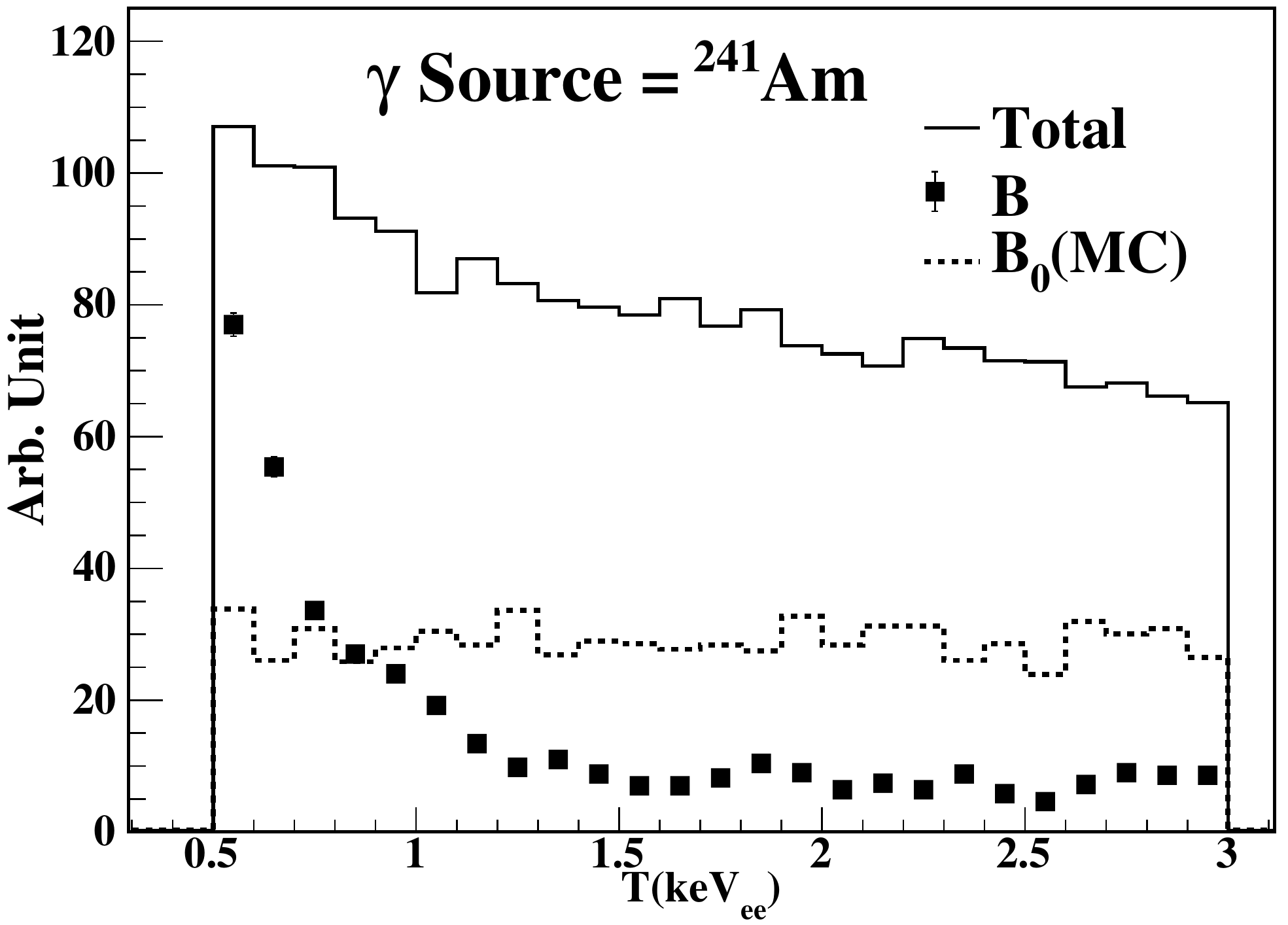}\\
{\bf (b)}\\
\includegraphics[width=7.5cm]{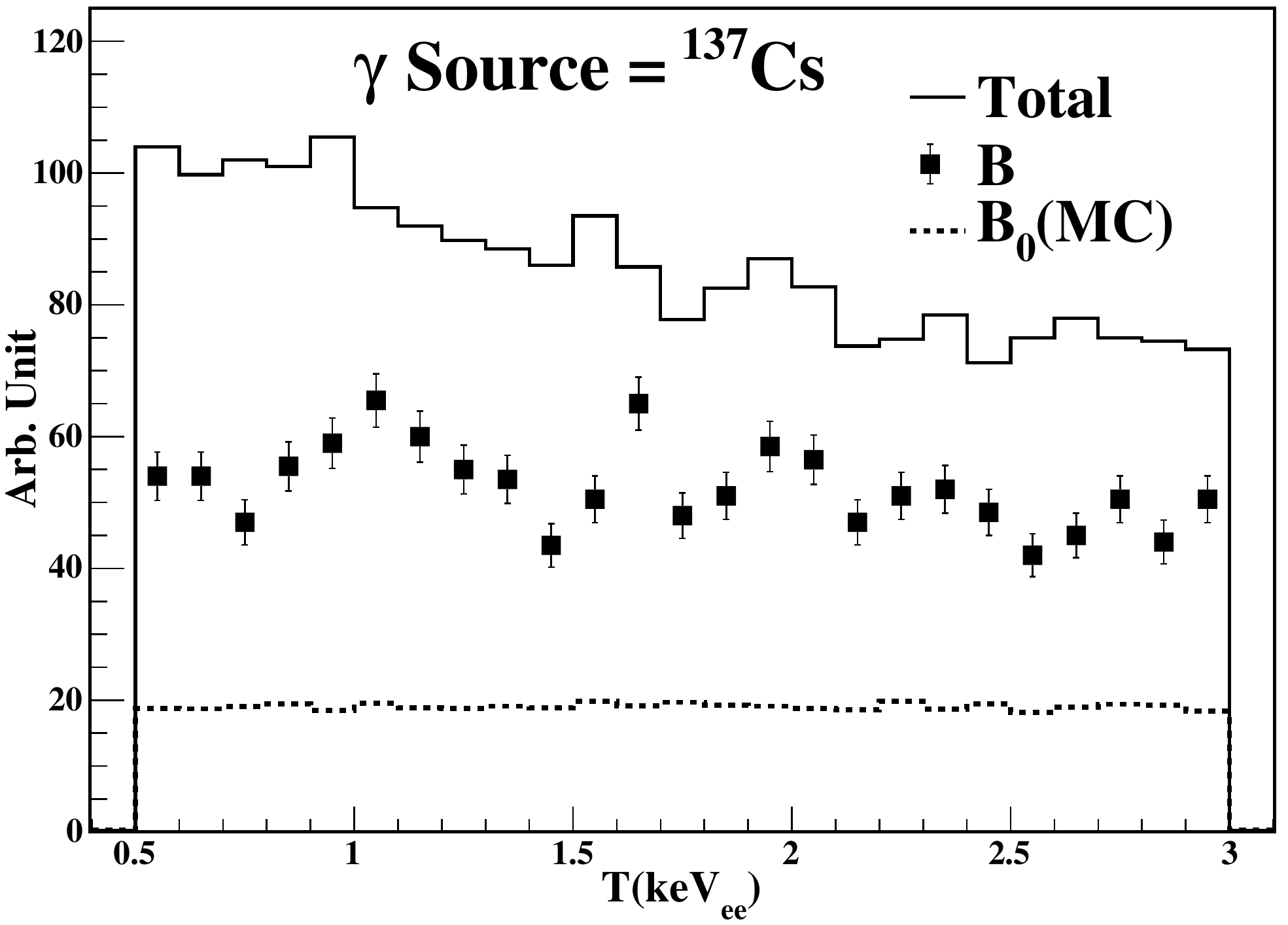}\\
{\bf (c)}\\
\includegraphics[width=7.5cm]{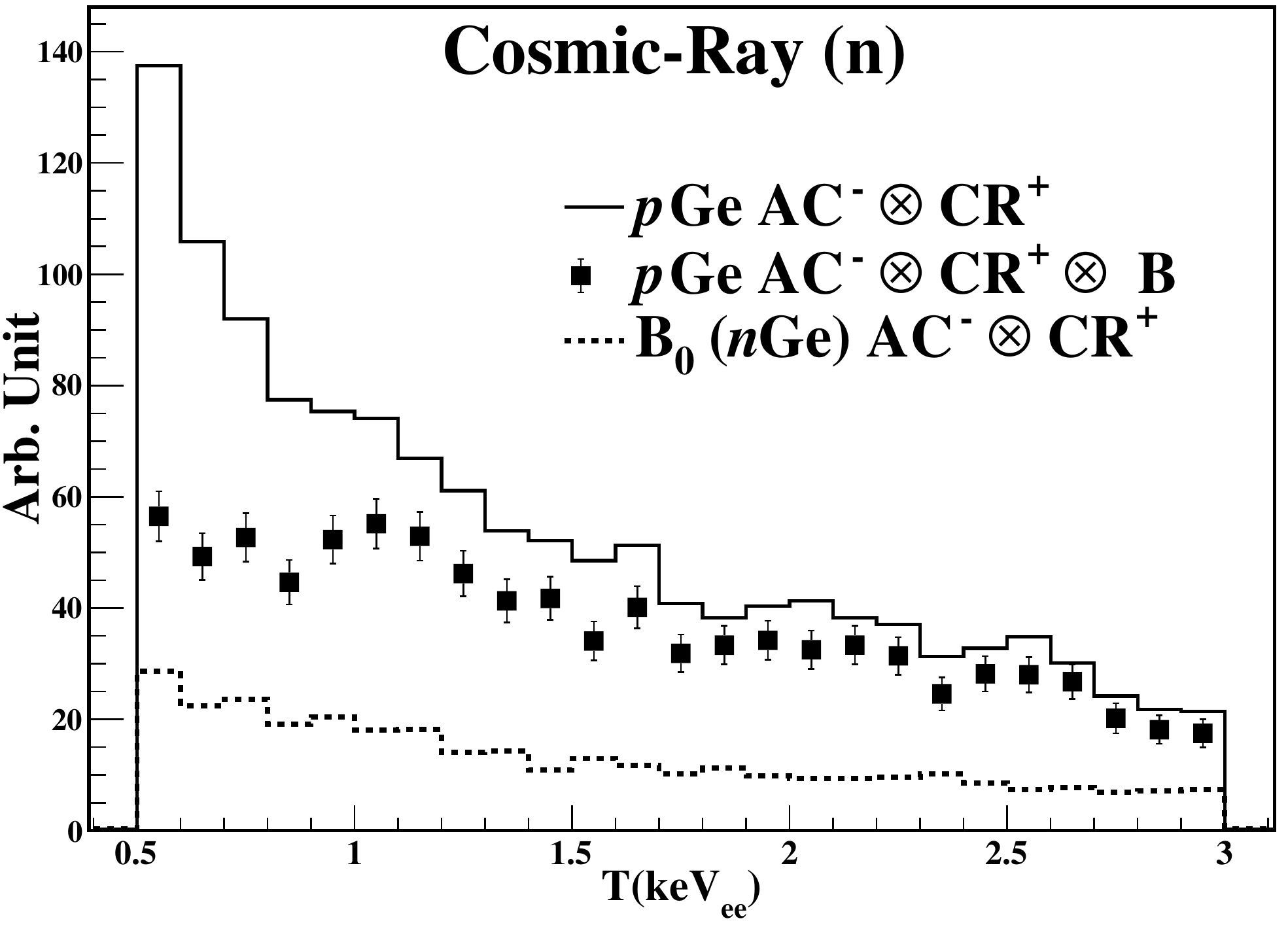}
\caption{
The derivation of
$( \effbs , \lmbdbs )$$-$
The measured Total and B spectra from $\pge$
with the
surface-rich
$\gamma$-ray $-$ (a) $^{241}$Am, (b)$^{137}$Cs,
as well as (c) the bulk-rich cosmic-ray induced neutrons.
They are compared to reference B-spectra
acquired through simulations for $\gamma$-rays
and $\nge$ measurement for cosmic-neutrons.
}
\label{fig::elcalib}
\end{figure}

\begin{figure}
{\bf (a)}\\
\includegraphics[width=8.5cm]{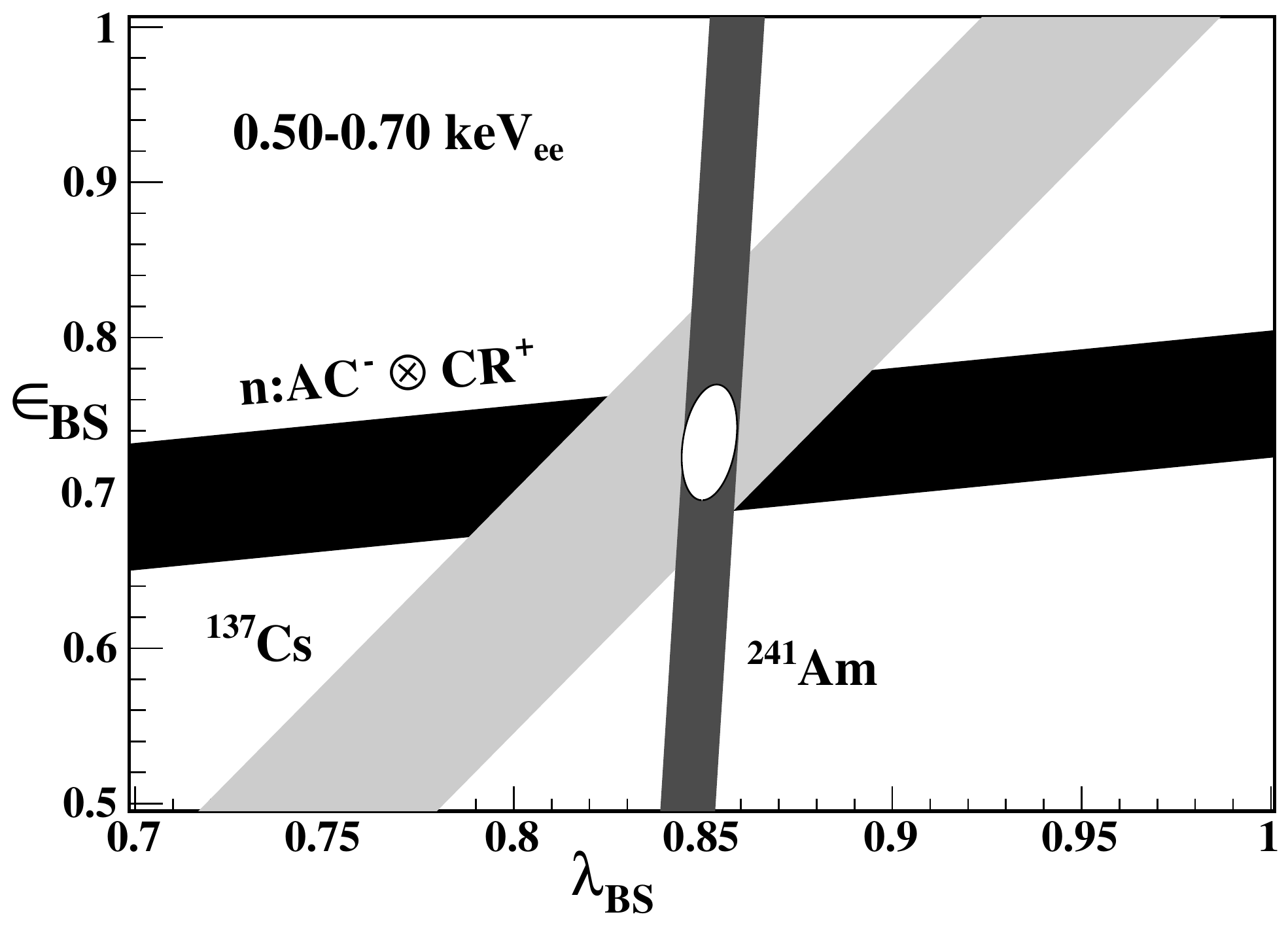}
{\bf (b)}\\
\includegraphics[width=8.5cm]{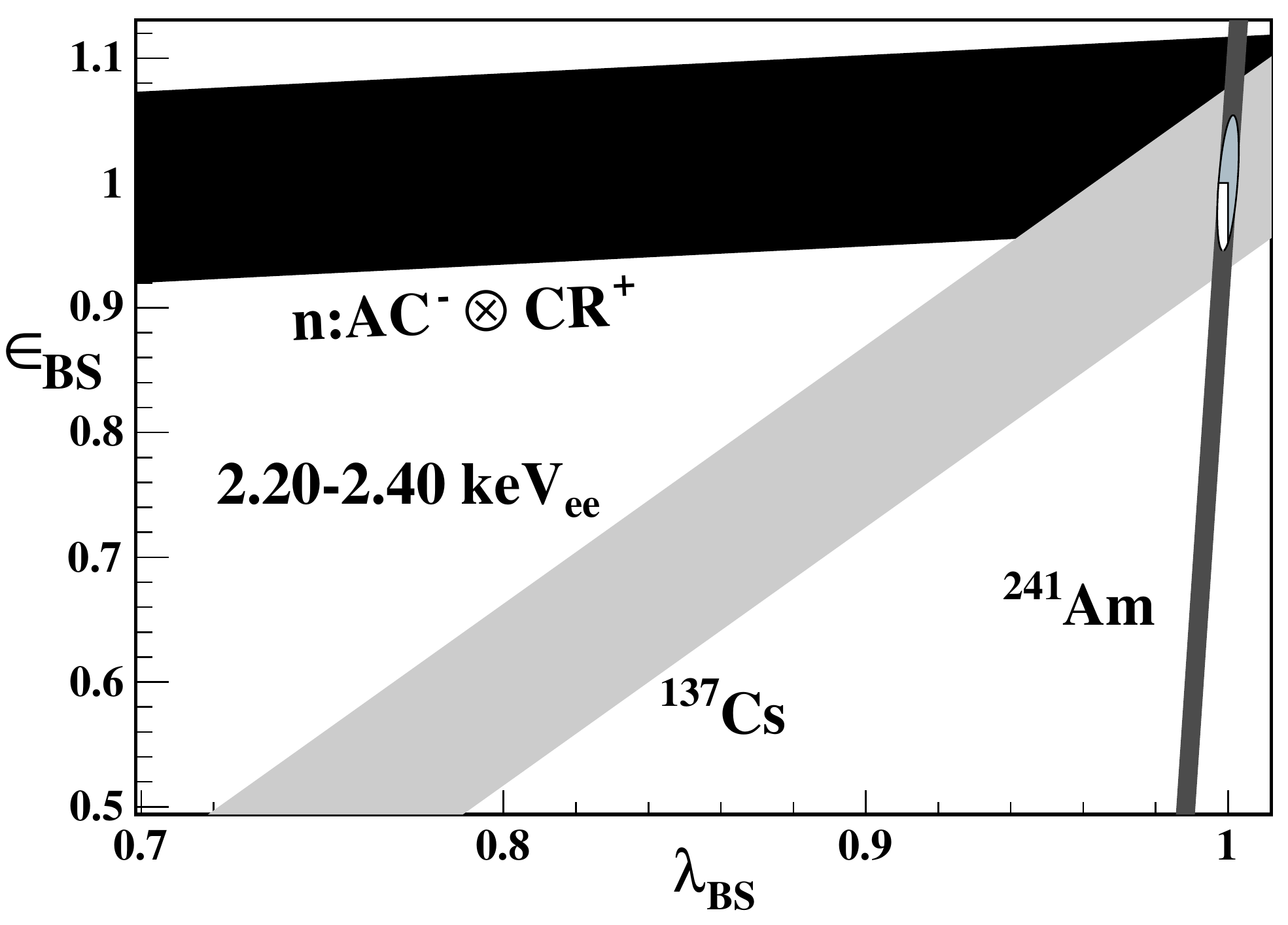}
\caption{
Allowed bands of
$( \effbs , \lmbdbs )$
derived by solving the coupled equations in
Eq.~\ref{eq::elcoupled} on the calibration data 
set, at (a) 0.5$-$0.7~$\keVee$, and
(b) an energy bin at 2.2~$\keVee$.
}
\label{fig::elbands}
\end{figure}

Anomalously slow rise-time events 
are observed in excess at low energy from
the AC$^{-}$$\otimes$CR$^{-}$ samples.
Their origin is not yet identified 
$-$ electronic noise is a possibility since
such events are uncorrelated with other detector
components.
These large-$\tau$ events are categorized as
``Discard'' in Figure~\ref{fig::bscut},
and are rejected in the subsequent analysis.
The $\sigma_{\tau}$ of the B-band at 500~$\eVee$
is 1.2~$\mu$s, such that
the mean is about 4.5~RMS from the 
Discard region.
The leakage of B-events is negligible, and  
there is no loss in signal efficiencies.

\begin{figure}
{\bf (a)}\\
\includegraphics[width=8.5cm]{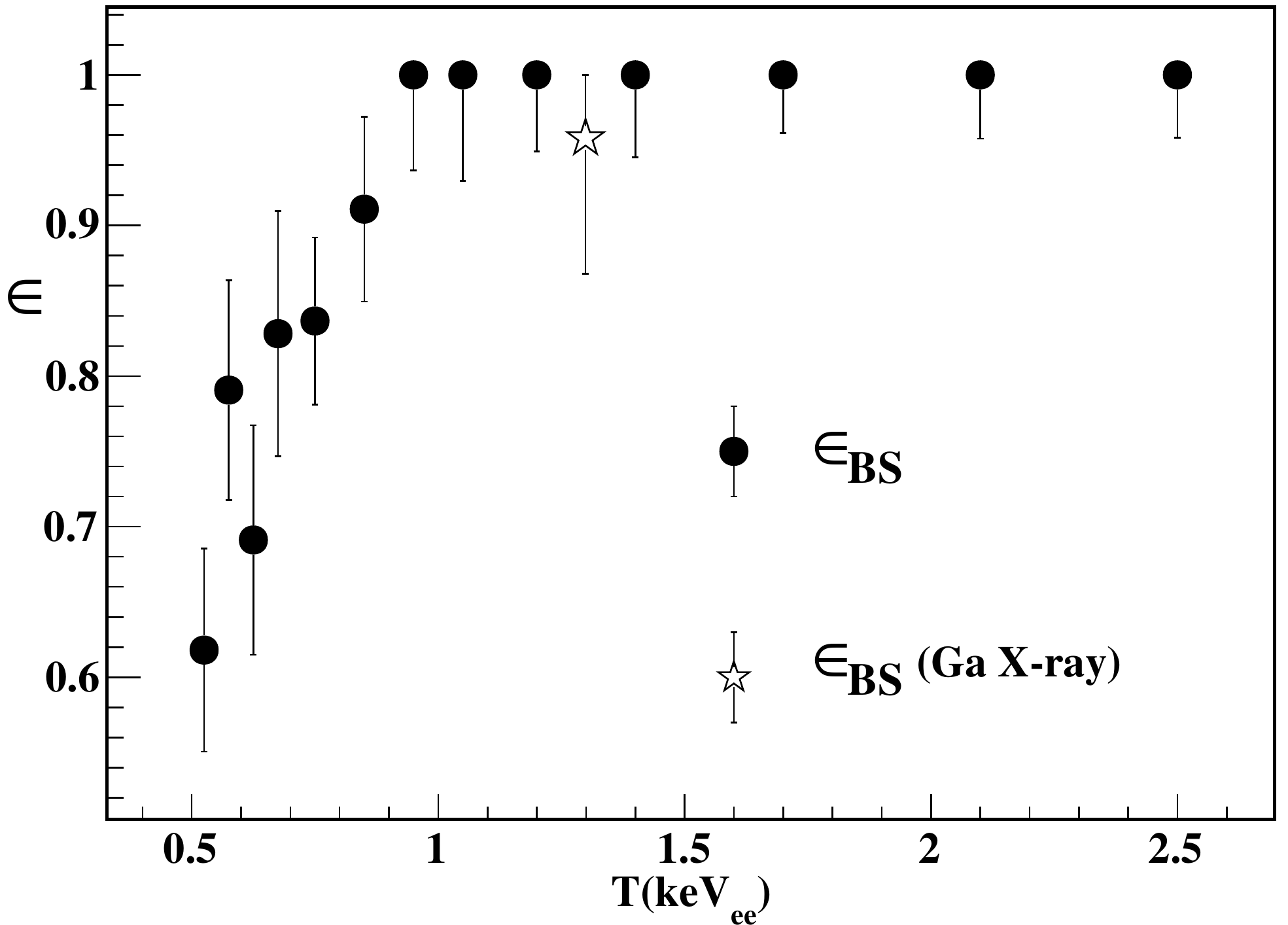}
{\bf (b)}\\
\includegraphics[width=8.5cm]{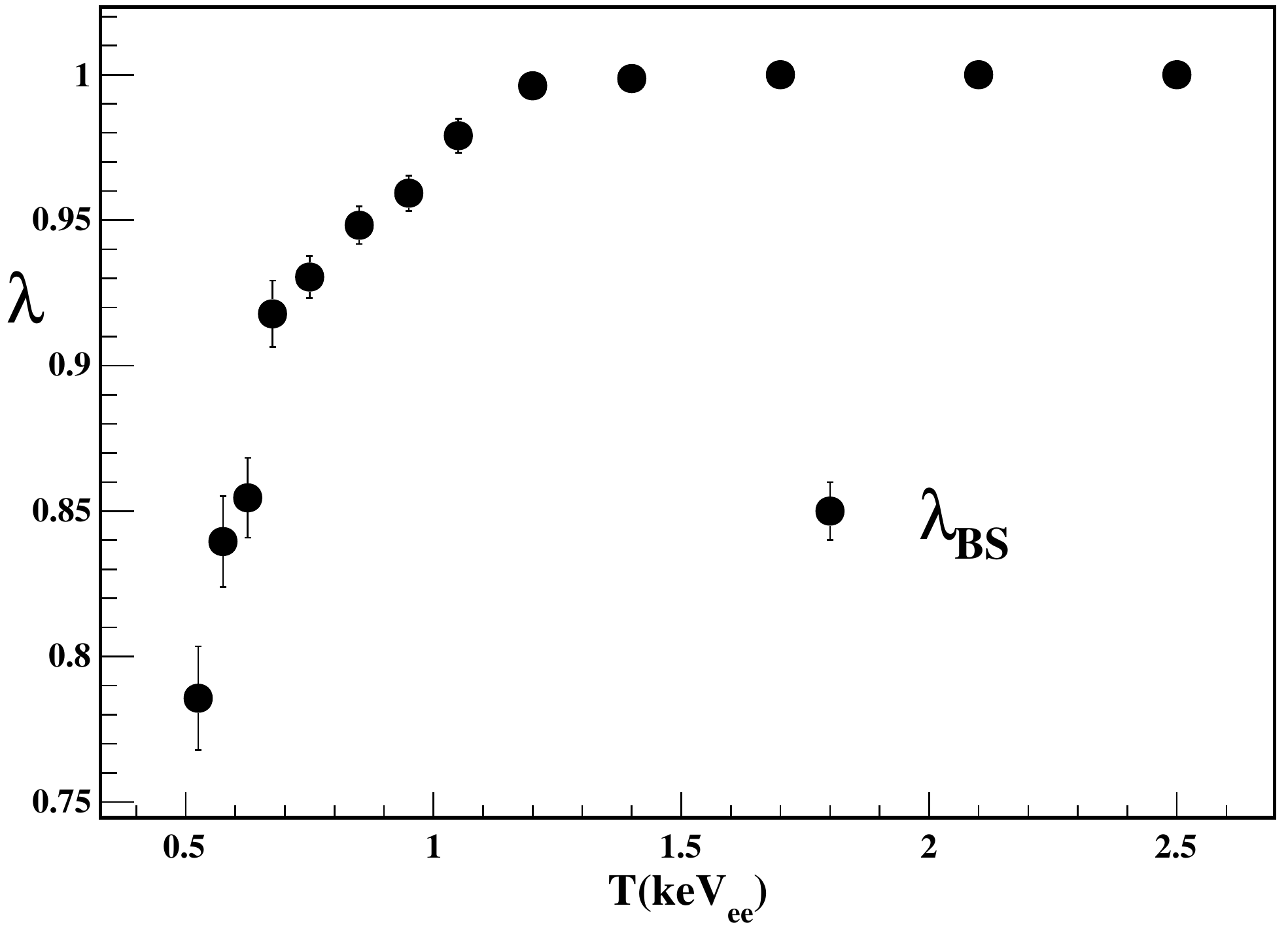}
\caption{
The measured (a) $\effbs$ and (b) $\lmbdbs$
as functions of energy.
Independent measurement on $\effbs$ with
Ga-L X-rays is included.
}
\label{fig::elresults}
\end{figure}


\section{Efficiencies Measurement and Correction}
\label{sect::bsel}

\subsection{Formulation}

Calibration of the BS-cut requires
the measurement of the 
bulk-signal retaining ($\effbs$) and 
surface-background suppressing ($\lmbdbs$)
efficiencies.
This is achieved by relating these 
efficiency factors with the observed and actual rates, 
denoted by (B,S) and ($\b0$,$\s0$), respectively.

The normalization assignment 
($\b0$,$\s0$)=(B,S) is made
on events within $T_0$=2.7-3.7~$\keVee$. 
It is equivalent to setting $\effbs$ and $\lmbdbs$ to 1.0.
This energy range is selected since it is above the
complications of the L-shell X-rays at $\sim$1~$\keVee$ 
as well as the physics region 
in dark matter analysis.

At lower energy, (B,S) and ($\b0$,$\s0$)
are related by the coupled equations:
\begin{eqnarray}
{\rm B} & = &  \effbs \cdot \b0 ~  
+  ~ ( 1 - \lmbdbs ) \cdot \s0 \nonumber \\
{\rm S} & = &  ( 1 - \effbs) \cdot \b0 ~  
+  ~ \lmbdbs \cdot \s0 ~~ , 
\label{eq::elcoupled}
\end{eqnarray}
with an additional unitarity constrain:
$\b0$+$\s0$=B+S.
The derivation of ($\effbs$,$\lmbdbs$)
therefore involves at least two
measurements of (B,S) where the
actual rates ($\b0$,$\s0$) are known.


\subsection{Calibration  Data}
\label{sect::calib}

The averaged TA and SA pulse shapes of
AC$^+$$\otimes$CR$^+$ physics samples,
together with random trigger and 
test pulser events, are displayed in 
Figures~\ref{fig::pulseshape}a\&b, respectively.

The pulser events exhibit 
different profiles as the physics samples,
and therefore are not appropriate for calibration purposes.
(We note, however, that the leading edge of
their shaped pulses are identical, such that pulser
events can be applied in the measurement of
trigger efficiencies.)
Calibration data with $^{241}$Am, $^{137}$Cs and {\it in situ}
cosmic-ray induced fast neutrons are adopted instead.
They have vastly different distributions in
their events locations within the detector
as illustrated in Figure~\ref{fig::eventdepth}a, 
and hence can play complementary roles in
the calibration.

\begin{enumerate}
\item 
{\bf Surface-rich events with $^{241}$Am and $^{137}$Cs  
$\gamma$-ray sources } $-$\\
As displayed in Figures~\ref{fig::elcalib}a\&b,
the measured B-spectra
are compared to the reference B
derived from full simulation
with surface layer thickness of 1.16~mm as input.
The simulated B-spectra due to external $\gamma$-sources
over a large range of energy
are flat for {\it T}$<$10~$\keVee$.\\
Consistent results are obtained for different source
orientations relative to the $\pge$ sensor.
The adopted data for analysis are those
with sources placed at the top facing the
flat surface of the $\pge$ crystal.
They provide the most accurate measurements since 
this is the direction
where the passive materials between the sources and
the crystal are minimal and their thickness is the
most uniform by construction.

\item
{\bf Bulk-rich events with cosmic-ray induced fast neutrons } $-$\\
A 523~g n-type point-contact
germanium ($\nge$) detector was constructed.
The components and dimensions are identical
to those of $\pge$.
The surface of $\nge$
is a p$^+$ boron implanted electrode of
sub-micron thickness.
There are no anomalous surface effects.
Data are taken under identical shielding configurations
at KSNL.
The trigger efficiency is 100\% above T=500~$\eVee$,
and energy calibration is obtained from the
standard internal X-ray lines.\\
The AC$^-$$\otimes$CR$^+$ condition selects
cosmic-ray induced fast neutron events
without associated $\gamma$-activities,
which manifest mostly as bulk events.
This tag in $\nge$ is therefore taken as 
the B-reference
and compared with those of
AC$^-$$\otimes$CR$^+$$\otimes$B in $\pge$,
as depicted in Figure~\ref{fig::elcalib}c.
\end{enumerate}

\subsection{Results on ($\effbs$,$\lmbdbs$)} 

Using the calibration data discussed above,
($\effbs$,$\lmbdbs$) are derived by
solving the coupled
equations in Eq.~\ref{eq::elcoupled}.
The three allowed bands at 0.5$-$0.7~$\keVee$ and at 
2.20$-$2.40~$\keVee$ are illustrated in 
Figures~\ref{fig::elbands}a\&b, respectively.
The different orientations of the bands are
consequences of different B:S ratios which
are due to the different 
penetration-depth distributions
of Figure~\ref{fig::eventdepth}a.
The surface-rich $\gamma$-events
and the bulk-rich cosmic-ray induced neutron-events
play complementary roles in constraining
$\lmbdbs$ and $\effbs$, respectively.
The bands have common overlap regions, indicating
the results are valid for the entire detector volume.
It is therefore justified to apply 
($\effbs$,$\lmbdbs$) derived from the calibration data
to the physics samples.

The energy dependence of ($\effbs$,$\lmbdbs$) 
are displayed in Figures~\ref{fig::elresults}a\&b.
By comparing the measured
{\it in situ} Ga-L X-ray peak at 1.3~$\keVee$
after BS-selection to that predicted by the corresponding
K-peak at 10.37~$\keVee$,
an additional data point on $\effbs$ 
is independently measured.
This provides a cross-check
to the calibration procedures and 
indicates consistent results.

The measured ($\effbs$,$\lmbdbs$) 
are close to unity at $T > 1.5 ~ \keVee$.
It follows from $\sigma_{m}$ being less than
the separation of the bands from $\tau_0$.
There is no leakage between the B- and S-events originated
away from the transition zone.
The ambiguity in the B/S assignment to events 
within the zone is accounted for 
through a systematic uncertainty
on the fiducial mass, which is negligible compared to
the total error from Table~\ref{tab::error}.


\begin{table*}
\begin{ruledtabular}
\begin{tabular}{lccc}
Energy Bin & 0.50$-$0.55~$\keVee$ & 0.95$-$1.00~$\keVee$
& 1.90$-$1.95~$\keVee$ \\ 
~~ Measurement and Total Error ($\cpkkd$) 
& 10.6$\pm$5.0 & 9.8$\pm$2.4 & 6.1$\pm$1.6 \\ \hline
\multicolumn{3}{l}{Relative Contributions to Total Error$^\dagger$~:}\\
I) Uncertainties on Calibration ($\effbs$,$\lmbdbs$) 
from Fig.~\ref{fig::elresults}~:
& 0.26 & 0.064 & $<$0.03 \\
\multicolumn{3}{l}{
II) Measurement Error on $\b0$ from Eq.~\ref{eq::b0s0}~:
} \\ 
~~~~~~~~Statistical Errors of (B,S) & 
\hspace*{1.5cm} \multirow{2}*{\{} 0.41 
& \hspace*{1.5cm} \multirow{2}*{\{} 0.90 
& \hspace*{1.5cm} \multirow{2}*{\{} 0.99 \\
~~~~~~~~Scaling by $1 / ( \effbs + \lmbdbs - 1 )$ & 
\hspace*{1.5cm} ~~2.29 
& \hspace*{1.5cm} ~~1.07 
& \hspace*{1.5cm} ~~1.00 \\
~~~~~~Combined & 0.95 & 0.96 & 0.99 \\
III) Systematic Uncertainties due to Parameter Choice~: & & & \\
~~~~~~~~(i) Rise-time Cut-Value $\tau_0$ & 
\hspace*{1.5cm} \multirow{5}*{\{} 0.12 
& \hspace*{1.5cm} \multirow{5}*{\{} 0.25 
& \hspace*{1.5cm} \multirow{5}*{\{} 0.09 \\
~~~~~~~~(ii) Fiducial Mass from Choice of $\tau_0$ & 
\hspace*{1.5cm} ~~0.03 & \hspace*{1.5cm} ~~0.06 & \hspace*{1.5cm} ~~0.06 \\
~~~~~~~~(iii) Normalization Range & 
\hspace*{1.5cm} ~~0.13 & \hspace*{1.5cm} ~~0.10 & \hspace*{1.5cm} ~~0.07 \\
~~~~~~~~(iv) ($\b0$,$\s0$)=(B,S) at Normalization & 
\hspace*{1.5cm} ~~0.08 &  \hspace*{1.5cm} ~~0.03 & \hspace*{1.5cm} ~~0.03 \\ 
~~~~~~~~(v) Choice of Discard Region & 
\hspace*{1.5cm} ~~0.05 &  \hspace*{1.5cm} ~~0.01 & \hspace*{1.5cm} ~~0.001 \\ 
~~~~~~Combined Systematic Error & 0.20 &  0.27 & 0.12 
\end{tabular}
\end{ruledtabular}
\caption{
Contributions to
the uncertainties on
the AC$^-$$\otimes$CR$^-$$\otimes$$\b0$
spectrum from various sources.
$^\dagger$ Errors are combined in quadrature. 
The total error is normalized to 1.0.
}
\label{tab::error}
\end{table*}


\section{Efficiencies-Corrected Background Spectra}

\subsection{Efficiencies-Corrected Spectra}

\begin{figure}
{\bf (a)}\\
\includegraphics[width=8.5cm]{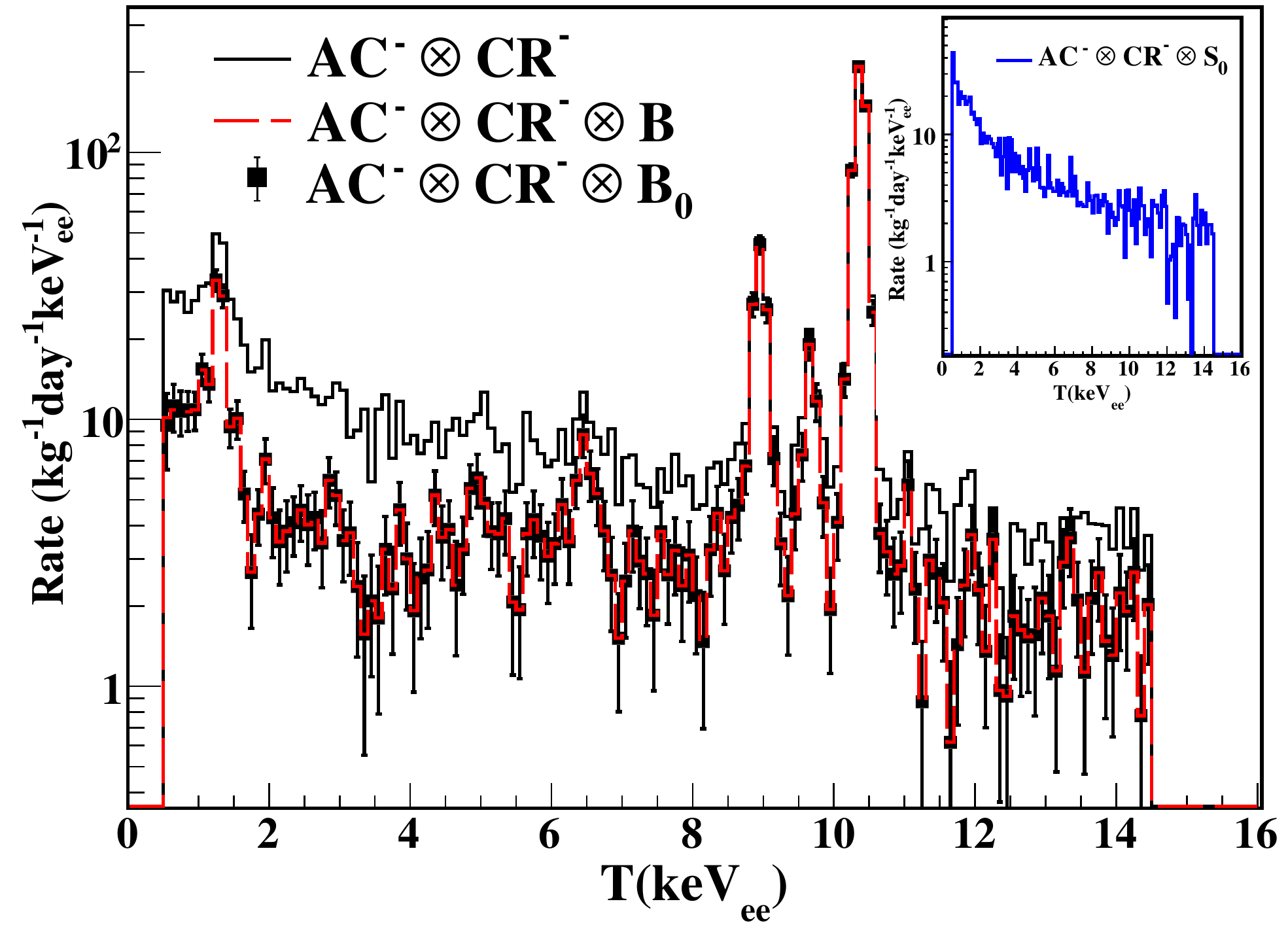}
{\bf (b)}\\
\includegraphics[width=8.5cm]{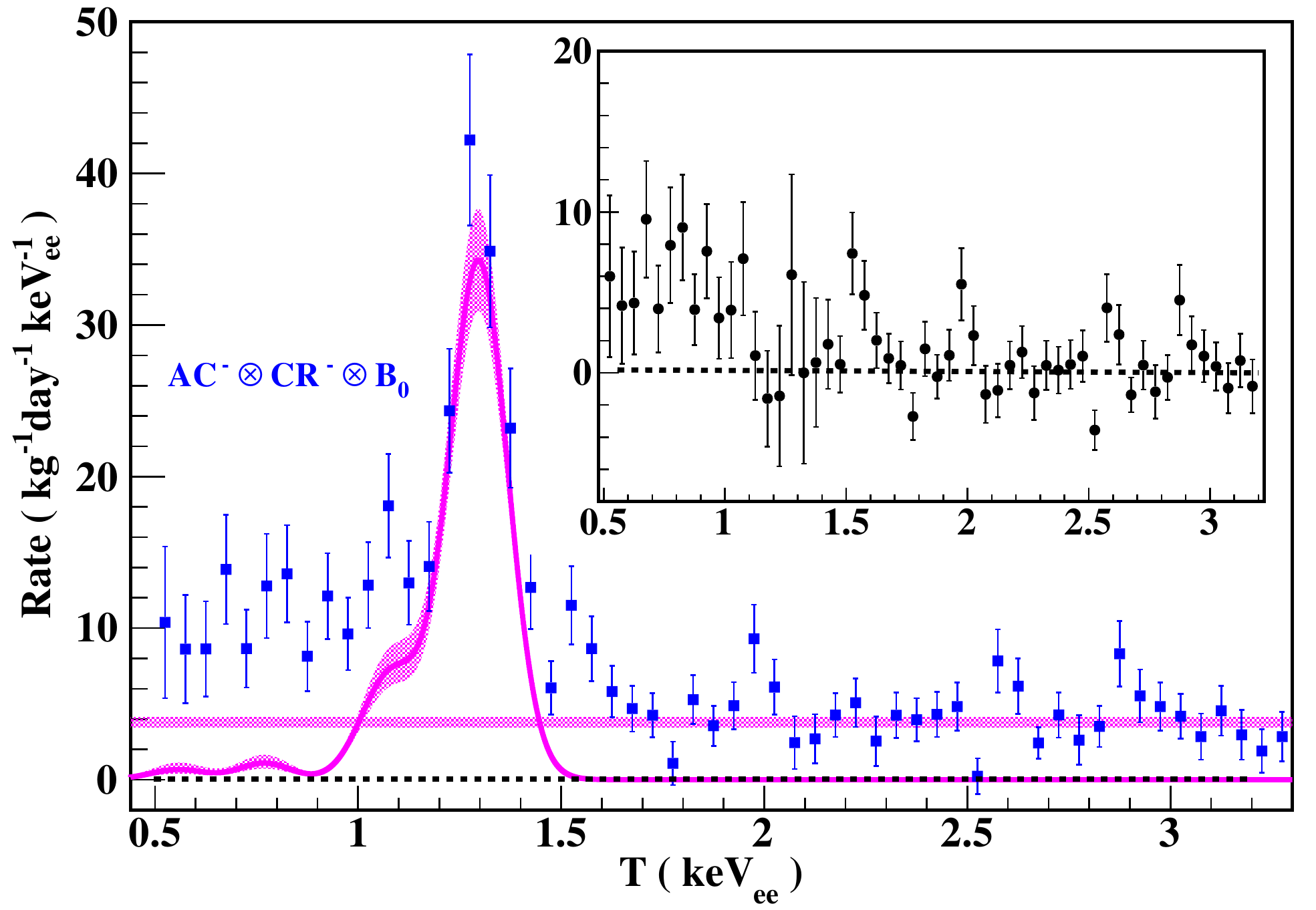}
\caption{
(a)
Measured and corrected spectra of
the AC$^-$$\otimes$CR$^-$ tag. 
(b)
Shown in magenta are flat background due to 
high-energy $\gamma$-rays
from ambient radioactivity, 
and contributions from the L-shell X-rays.
Depicted in inset is 
the residual spectrum after background subtraction,  
corresponding to candidate $(\chi / \nu ) N$ events. 
}
\label{fig::b0vv}
\end{figure}

Once $( \effbs , \lmbdbs )$ are measured
with the calibration data,
the efficiency-corrected ($\b0$,$\s0$)
of the physics samples can be derived via
the solution of Eq.~\ref{eq::elcoupled}:
\begin{eqnarray}
\b0 & = & 
\frac{ \lmbdbs \cdot {\rm B } - ( 1 - \lmbdbs ) \cdot {\rm S } }
{ ( \effbs + \lmbdbs - 1 ) } \nonumber \\
\s0 & = & 
\frac{ \effbs \cdot {\rm S } - ( 1 - \effbs ) \cdot {\rm B } }
{ ( \effbs + \lmbdbs - 1 ) } ~~ . 
\label{eq::b0s0}
\end{eqnarray}
The formulae can be understood as follows:
$\b0$($\s0$) should account for
the loss of efficiency in the measurement of B(S) 
in the first positive term,
followed by a subtraction of the leakage effect from S(B)
in the second negative term.

The AC$^-$$\otimes$CR$^-$ tagged events
from $\pge$ data taken at KSNL 
at various stages of the analysis
are depicted in Figure~\ref{fig::b0vv}a.
The measured-B and corrected-$\b0$ spectra
are almost identical. 
At $T > 1.5~ \keVee$, this is a direct consequence
of $\effbs = \lmbdbs = 1$. 
At low energy, the efficiency-correcting and 
background-subtracting effects
compensate each other in this data set.

After subtracting a flat background due to
high energy $\gamma$-rays and the known L-shell
X-rays contributions predicted by the
accurately-measured K-peaks at higher energy, the
residual spectrum is shown in the inset 
of Figure~\ref{fig::b0vv}b.
It still shows excess of events
at the sub-keV region. 
The origin is not yet identified,
and the studies towards the understanding of which 
are intensely pursued.
Under the conservative assumption that WIMPs signals
cannot be larger than the residual excess, 
constraints on $\chi N$ cross-section
versus $\mwimp$ were derived. 
They probed and excluded some of the allowed
regions on light WIMPs from 
earlier experiments~\cite{texono2013}.

\subsection{Error Sources and Assignment}

The errors on ($\effbs$,$\lmbdbs$) are
shown in Figures~\ref{fig::elresults}a\&b.
They are derived from the global fits on
the allowed bands in Figure~\ref{fig::elbands}.
Standard error propagation techniques are applied
to derive the resulting uncertainties 
on ($\b0$,$\s0$) via Eq.~\ref{eq::b0s0}.

The uncertainties include contributions from 
their own measurement errors, 
the ($\effbs$,$\lmbdbs$) calibration errors, 
as well as systematic uncertainties.
Their relative contributions in three representative
energy bins are summarized in Table~\ref{tab::error}.
The leading contribution is 
the statistical errors on (B,S), 
scaled by a factor 1/$( \effbs + \lmbdbs - 1 )$.
This can be seen from the structure of the
formulae in Eq.~\ref{eq::b0s0}.
The total errors therefore increase  
as $\effbs$ and $\lmbdbs$ deviate from unity 
towards the analysis threshold of 500~$\eVee$.

Systematic uncertainties on the BS-selection
procedure originate from the choice of $\tau_0$
and its effect on the $\pge$ fiducial mass,
the choice of the normalization energy 
range at $T_0$,
the assignment of ($\b0$,$\s0$)=(B,S) 
in this interval,
as well as the choice of the Discard region.
They are estimated by the shifts in $\b0$
as the parameters are varied in the vicinity
of their optimal values. 
As an illustrated example,
the ``$\tau$-scan'' range for $\tau_0$ 
is depicted in Figure~\ref{fig::bscut}a.
In all cases, the shifts are small 
compared to the total errors.
Accordingly, the contributions of systematic uncertainties 
are minor, as illustrated in Table~\ref{tab::error}.


\section{Conclusion and Prospects}

The results on ($\effbs$,$\lmbdbs$) calibration 
and the subsequent ($\b0$,$\s0$) measurements in $\pge$ 
confirm that both signal efficiencies and background leakage 
to the signal region are crucial in the analysis, 
all the more so since the efficiency factors are significantly 
less than unity at the analysis threshold,
which is 500~$\eVee$ in this work.
A mis-placement of $\lmbdbs$=1 would introduce excess of events 
at low energy which could have been mis-interpreted as 
signatures of light WIMPs. 
Conversely, genuine WIMP signals can also be masked out
through an incorrect assignment of the factors.

We note that it is necessary to derive ($\effbs$,$\lmbdbs$) 
and ($\b0$,$\s0$) by solving 
the coupled equations Eq.~\ref{eq::elcoupled}
to obtain Eq.~\ref{eq::b0s0}. 
If the two efficiency corrections
were performed separately, an incorrect expression of 
$\b0 = {\rm B} / \effbs -  [ ( 1 - \lmbdbs ) \cdot {\rm S} / \lmbdbs ]$
would follow.
Deviations from the correct values
would depend on the B:S ratio, and would
be more pronounced when ($\effbs$,$\lmbdbs$) decrease.
A relative error of order unity would be introduced
to $\b0$ for this data set at threshold.
We note that this comment as well as 
Eqs.~\ref{eq::elcoupled}\&\ref{eq::b0s0} 
also apply to generic
event selection procedures 
in cut-based analysis.

Despite advances in the BS-selection and efficiency factors
measurements discussed in this report, 
there are still fundamental challenges to further boost 
the sensitivities on the studies of sub-keV events with $\pge$.
The 1/$( \effbs + \lmbdbs - 1 )$ factor of 
Eq.~\ref{eq::b0s0} increases the uncertainties 
to the physics signal $\b0$ near threshold.
In addition, $\b0$ depends on measurements of all of
the input parameters ($\effbs$,$\lmbdbs$,B,S). 
This calls for caution in the 
investigations of time variation and modulation effects on $\b0$, 
in which the time stabilities of these input have to 
be independently monitored.

To overcome these difficulties, the elimination of the
anomalous surface effects at the hardware raw signal level
in Ge detectors is much more desirable. To these ends,
the merits and operation of $\nge$ are being studied. 
This detector has already proved crucial
to provide calibration data to the $\pge$. 
Research efforts are being pursued to turn it 
into a target with comparable sensitivities.

A by-product of this investigation is that 
the $\pge$ $\tau$ distributions are different
for different sources, as depicted in 
Figure~\ref{fig::bscut}b. 
Therefore, the studies of signal rise-time
may shed light on the nature of the background.
Further systematic and quantitative studies are 
under way.

\section{Acknowledgment}

This work is supported by
the Academia Sinica Investigator Award 2011-15,
contracts 99-2112-M-001-017-MY3 and
102-2112-M-001-018
from the National Science Council, Taiwan.

\end{document}